\documentclass{article}
\usepackage{spconf,amsmath,graphicx}
\usepackage{lgrind}
\usepackage{cmap}
\usepackage{booktabs}
\usepackage{floatrow}
\usepackage{lipsum}
\usepackage{amsfonts}
\usepackage{mathtools}
\usepackage{algorithm}
\usepackage{algpseudocode}
\usepackage{multirow}
\usepackage{csquotes}
\usepackage{rotating}
\usepackage{longtable}
\def\L{{\cal L}}
\title{Progressive distillation diffusion for raw music generation}
\name{Svetlana Pavlova}
\address{Skolkovo Institute of Science and Technology, 143026 Moscow, Russia}
\begin{document}
%
\maketitle
\begin{abstract}
This paper aims to apply a new deep learning approach to the task of generating raw audio files. It
is based on diffusion models, a recent type of deep generative model.
This new type of method has recently shown outstanding results with image generation. A lot of focus
has been given to those models by the computer vision community. On the other hand, really few have
been given for other types of applications such as music generation in waveform domain.

In this paper the model for unconditional generating applied to
music is implemented: Progressive distillation diffusion with 1D U-Net. Then, a comparison of different parameters of diffusion and their value in a full result is presented. One big advantage of the methods implemented
through this work is the fact that the model is able to deal with progressing audio processing and generating , using transformation from 1-channel 128 x 384 to 3-channel 128 x 128 mel-spectrograms and looped generation. The empirical comparisons are realized across  different self-collected datasets.
\end{abstract}
\section{Introduction}
\label{sec:intro}

\subsection{Relevance}
The expansion of soft computing technology has opened up new opportunities for artificial intelligence, particularly neural networks, to tackle a variety of problems. Although the classification and generation of images using neural networks have garnered widespread popularity, their applicability to other tasks remains uncertain. One such area is the processing of audio signals through artificial intelligence. The research in this area has started with the identification and classification of acoustic signals and has progressed to the study of generating new audio signals, particularly music. This has become an ideal domain for study similar to that of image generation.

The difficulty of such a work stems from the fact that music is a sequential data set in which the set of information, consonance harmony, overall rhythm, and timbre are all crucial. Musical compositions, in general, include many instruments or voices; it is important to construct these sound sequences so that they can merge.

While the methods used to analyze the musical range will advance in the fields of processing and analyzing audio signals, frequency spectra, the solution to such a problem can find its niche in the imposed application for sound design, develop the route of performers in the field of science and art.

In classical music, when the composer edits the stream solely by ear, the issue might not be as clear, but as sound engineering has advanced, it has been much simpler to employ analyzers to eliminate noises like ultrasound, for instance. There are already numerous efforts for creating music using artificial intelligence, but there is still need for study in the area of raising the standard of generative music so that it is as near to human-generated as feasible.

There has been an increase in interest in employing neural networks to create music as the usage of deep learning methods has become more widespread in recent years. Neural networks are used in deep generative models to simulate high-dimensional probability distributions. In particular, for the purposes of this work, we will concentrate on the application of the denoising diffusion probabilistic model, a novel and promising deep generative modeling technique. These techniques have produced amazing results in the fields of computer vision and natural language processing, but they are rarely used to create music in the raw wave domain.

The diffusion model excels in music production due to its ability to detect intricate connections and patterns in the data. However, the input representation and computing capabilities must be carefully taken into account when employing the diffusion model for music production. It has been demonstrated that spectrograms, which are visual representations of the frequencies and amplitudes of sound sources, are a useful input representation for diffusion model-based music creation.

\subsection{Literature review} A preliminary literature review shows that past studies are primarily focused on using popular architectures to generate notes primarily in the classical music genre. In terms of modeling various approaches have been recommended.

The authors of the article~\cite{pasini2022musika} propose an improved GAN-based model for generating raw music. The initial data is passed into the autoencoder, which converts the sequence into low-dimensional latent vectors. Next, the vectors are loaded into the GAN (adapted FastGAN architecture), within which new sequences are produced utilizing style conditioning vector. At last, the generated latent vectors are fed into decoder, to be reconverted into a waveform. The shift to and usage of latent representations of audio is a feature that increases the model's speed. The model generally strikes a fair balance between the accuracy of the generated data and the bandwidth of training and generation, making it fascinating to investigate.

The authors of the article~\cite{mehri2016samplernn} propose a SampleRNN model that uses a hierarchical architecture to model dependencies in audio data. Data input is done via raw waveforms. It is made easier to learn the data directly from audio samples because a hierarchy of time scales and frequent updates are employed to solve the difficulty of representing incredibly high temporal data. The model has various modules, including: a deep RNN module called a "frame-level module" that condenses the history of its inputs into a conditioning vector for the following module down, A multilayer perceptron with the Softmax function is the sample-level module (a q-way Softmax is an output of the MLP). With each sequence broken into smaller subsequences (of length 512), truncated backpropagation through time is employed for training, with gradients propagated solely to the beginning of each subsequence.

The authors of the article~\cite{oord2016wavenet} propose a WaveNet model based on the PixelCNN architecture for generating raw audio waveforms. The model consists of casual convolutional layer, several stacks of dilated conv layers and gated blocks, skip connections, relu, 1x1 convolution, relu, 1x1 convolution and softmax with final mulaw quantization for mapping the output values. Additionally, the model can be conditioned on other input variable with global or local conditioning.

Another solution in a waveform domain is~\cite{chen2020wavegrad}. The proposed model is WaveGrad, a conditional model for waveform generation which estimates gradients of the data density. WaveGrad uses a gradient-based sampler conditioned on the mel-spectrogram to repeatedly update the signal starting with Gaussian white noise. The model implicitly optimizes for the weighted variational lower-bound of the log-likelihood, needs only a constant number of generation steps during inference, and is non-autoregressive.

In the paper~\cite{liu2020unconditional}, the (GAN)-based model for unconditional generation of the mel-spectrograms of singing voices is proposed. The model is built using a boundary-equilibrium GAN architecture, which accepts various noise vectors as input. The temporal coherence of the produced samples is controlled by the generator using a hierarchical structure. By establishing loop consistency between each input noise vector and the corresponding segment in the output mel spectrogram, the learning process of the model is regulated.

The diffusion model for wave domain is proposed by~\cite{kong2020diffwave}. The authors suggest DiffWave, a flexible diffusion probabilistic model for generating waveforms with and without conditions. It easily reproduces the subtleties of waveforms conditioned on mel-spectrograms for the neural vocoding challenge and equals the strong autoregressive neural vocoder in terms of speech quality. It accurately captures the significant variances present in the data in both unconditional and class-conditional generation tasks, resulting in believable voices and consistent word-level pronunciations.

Another approach is demonstrated in paper~\cite{schneider2023archisound}. The authors suggest ArchiSound, a combination of tools and models for generating and manipulating music with diffusion. The models handle tasks including upsampling, vocoding, latent text-to-audio creation, and unconditional and conditional music generation. They train and create samples for the unconditional generation problem using the v-objective diffusion approach. The authors employ stacked one-dimensional UNets, which are inspired by the latent diffusion, as a technique in latent generation to produce text-to-audio quickly and precisely.

 The authors of the paper~\cite{huang2023noise2music} come up with a Noise2Music model for text-to-audio generation. Two different sorts of diffusion models are put forth for generation: a generator model that creates an intermediate representation that is text-conditioned and a cascade model that creates high-quality sounds that are trained and utilized consistently to create high-quality music. The research examines two possibilities for intermediate representation, one utilizing a spectrogram and the other with lower fidelity audio.
The fundamental claim is that the produced audio extends beyond the text prompt's fine-grained semantics and is also capable of properly portraying the text prompt's essential aspects, such as genre, pace, instruments, mood, and era.

 One more text-to-audio solution is AudioLDM, proposed by~\cite{liu2023audioldm}.
 The authors emphasize higher generation quality with less expensive computing. The suggested model is a text-to-audio system that uses tools for latent language and sound preconditioning (CLAP) to learn continuous sound representations. Text embedding may be used as a condition while sampling in pre-trained CLAP models to train LDM with audio embedding.
AudioLDM offers the benefit of both high generation quality and computational efficiency since it learns hidden representations of audio signals and the components that make them up without modeling cross-modal interactions. Trained on AudioCaps using a single GPU, AudioLDM achieves TTA's performance, which is assessed using both objective and arbitrary metrics (for example, Frechet distance). Moreover, AudioLDM is the first TTA system to provide a variety of text-based audio alterations in zero-shot mode, including style transfer.

In the article~\cite{leng2022binauralgrad}, BinauralGrad model is proposed for binaural sound synthesis from monophonic audio. In addition to the fundamental physical distortion of mono audio, this synthesis technique additionally applies head/ear filtering and room reverb. The authors approach the synthesis process from a different angle, decomposing the binaural sound into two distinct parts: a common element that is shared by the left and right channels and a particular part that is unique to each channel. The proposed model is described as a novel two-stage structure with diffusion models for each of their synthesis.
Because of monophonic sound, generic information about binaural sound is obtained using a single-channel propagation model in the first stage. Then, using a two-channel propagation model in the second stage, binaural sound is produced. The suggested BinauralGrad is capable of producing precise and excellent binaural audio samples by combining this unique viewpoint of two-staged synthesis with diffusion models.

The authors of the article~\cite{han2022nu} propose a NU-Wave 2 model for upsampling task. Using a single model, the moodel enables the generation of 48 kHz audio streams from inputs with various sampling rates. Based on the NU-Wave architecture, NU-Wave 2 employs Bandwidth Spectral Transformation (BSFT) to match input bandwidths in the frequency domain and Short-Time Fourier Convolution (STFC) to produce harmonics.

Because of the high expense of computing, there are a lot of contributions to processing music as syllables but predictably less solutions for wave domain. It is still appropriate to utilize RNN-based architectures for the task of generating music, although interest in diffusion models has grown recently as a result of their successful application to the task of generating visuals. The majority of articles on diffusion show the variety of ways that diffusion models may be used to create and process music. Diffusion models provide a potent tool for comprehending and modifying musical sounds, from music analysis and creation to audio signal processing and performance analysis.

\subsection{Individual contribution}
Many areas of research have been identified in the process. Firstly, the raw audio unconditional generation has seldom seen diffusion. Furthermore, the majority of the research that have been done so far on the application of generative models in music do not include further preprocessing or high-resolution mel-spectrograms with cutting-edge diffusion enhancements.
This paper primary contributions are:
\begin{enumerate}
    \item For the scientific community, provide implementations of a new deep generative methods able to handle high-resolution mel-spectrograms to generate music.
    \item For the deep learning community, show another application where diffusion can compete with state-of-the-art deep generative models such as GANs, VAEs, and others.
\item Demonstrate the competitiveness of proposed model.
\end{enumerate}

\begin{figure*}[h!]
    \centering
\includegraphics[width=\linewidth]{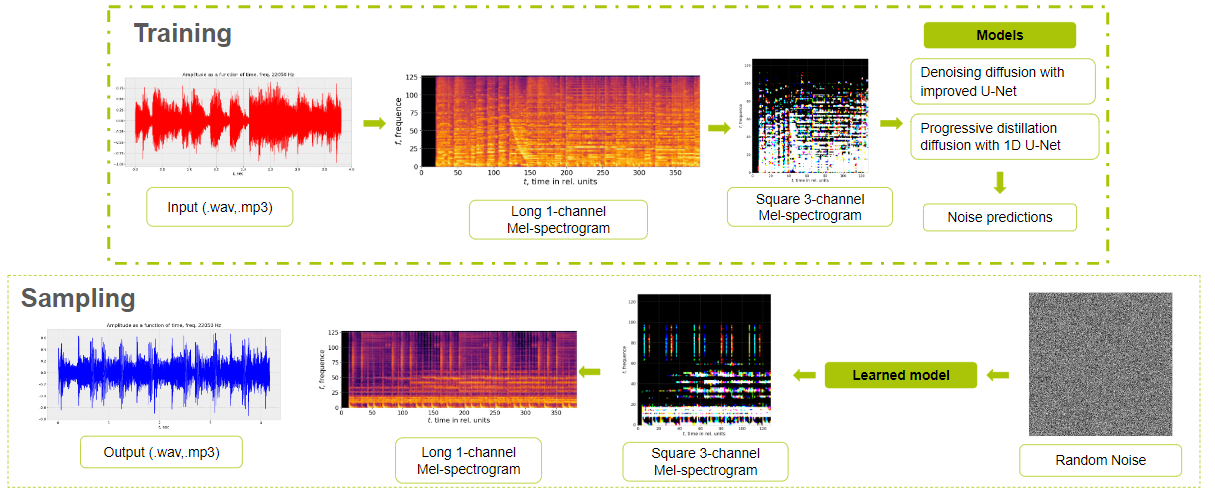}
    \caption{Main idea of the method} 
    \label{pic:scheme}
\end{figure*}

Up to this point, the field of music processing and generation was studied together with different approaches to initial data representation.
1D U-Net model with time embeddings and attention improvement and progressive distillation diffusion was implemented for denoising diffusion process, the precise idea can be seen in figure \ref{pic:scheme}. And finally, additional pre-processing and post-processing approaches for generating raw audios were presented.

\section{Preliminaries}
\label{sec:format}

\subsection{Audio Generation}
\label{ssec:subhead}

In recent years, there has also been an increase in interest in the unconditional creation of raw audio waveforms or its time-frequency representations. The Zero Resource Speech Challenge (also known as TTS without T or text-to-speech without text), which is held in several editions of the INTERSPEECH conference~\cite{dunbar2019zero}, is a famous example. The participants must unsupervisedly identify the subword units of speech in a collection of raw audio without text or phoneme labels in order to synthesis new utterances from new speakers.
The job setting is mostly speech-specific, therefore such TTS without T models are not easily transferable to create other sorts of audio signals, such as instrumental or ambient sounds, even if it is a fascinating and relevant unconditional audio production challenge.
 
\subsection{Diffusion}
\label{ssec:subhead}

The primary developed model is shown in this section: Denoising diffusion probabilistic models (DDPM). A group of latent variable models are called DDPMs. They were originally made available in 2015. They have gradually gained popularity ever since.
Subsequently, they demonstrated that they could complete the task of image synthesis with scores and performance that were competitive. A modified version of the DDPM model was also presented forth, and it has recently remained widely accepted. With a modified DDPM that proved successful in outperforming generative adversarial networks for the job of picture production, DDPMs have become quite popular. Although GANs have long considered of state-of-the-art for image generation, this represents a significant advancement.

\subsubsection{Forward diffusion process}

Given a data point sampled from a real data distribution~$x_0\sim q(x)$, let define a~forward diffusion process~in which a small amount of Gaussian noise is added to the sample in~$T$~steps, producing a sequence of noisy samples~$\{x_1,\,\ldots,\,x_T\}$. The step sizes are controlled by a variance schedule~${\{{\beta }_t\in (0,1)\}}^T_{t=1}$.
\[q\left(x_t\mathrel{\left|\vphantom{x_t x_{t-1}}\right.\kern-\nulldelimiterspace}x_{t-1}\right)=\mathcal{N}\left(x_t;{\mu }_t=\sqrt{1-{\beta }_t}\ x_{t-1},{{\sigma }_t=\beta }_tI\right)\] 
\[q(x_{1:T}|x_0)=\prod^T_{t=1}{q\left(x_t\right|x_{t-1})}\] 

Here $I$~is the identity matrix, indicating that each dimension has the same standard deviation~${\beta }_t{}$. Note that~$q\left(x_t\mathrel{\left|\vphantom{x_t x_{t-1}}\right.\kern-\nulldelimiterspace}x_{t-1}\right)$~is still a normal distribution, defined by the mean~$\mu $~and the variance~$\sigma $~where~${\mu }_t=\sqrt{1-{\beta }_t}\ x_{t-1}${} and~${{\sigma }_t=\beta }_tI$~.\textit{ }$\sigma $ will always be a diagonal matrix of variances (here~${\beta }_t${})

The data sample~$x_0$~progressively loses its distinctive characteristics as the step t increases. Eventually when~$T\to \infty ,~x_T$~is equivalent to an isotropic Gaussian distribution.

\subsubsection{Reverse diffusion process}

The reverse process iteratively refines a Gaussian noise input,~$x_T\sim N(0,I)$ into data point $x_0$. Unfortunately, $q(x_{t-1}|x_t)$~cannot easily estimated, because it needs to use the entire dataset and therefore, it is needed to learn a model~$p_{\theta }$~to approximate these conditional probabilities in order to run the~reverse diffusion process. If~${\beta }_t$~is small enough,~$q(x_{t-1}|x_t)$,~will also be Gaussian and $p_{\theta }$ can be chosen to be Gaussian and just parameterize the mean and variance.
\[p_{\theta }\left(x_{0:T}\right)=p(x_T)\prod^T_{t=1}{p_{\theta }{(x}_{t-1}|x_t)}\] 
\[p_{\theta }{(x}_{t-1}|x_t)=\mathcal{N}\left(x_{t-1};{\mu }_{\theta }\left(x_t,t\right),{\sigma }_{\theta }(x_t,t)\right)\] 

By additionally conditioning the model on timestep~$t$, it will learn to predict the Gaussian parameters ${\mu }_{\theta }\left(x_t,t\right)\ $and the covariance matrix~ ${\sigma }_{\theta }\left(x_t,t\right)$ for each timestep.

Thus, with those two processes, if one take $x_t$ and $x_{t-1}$, the forward process starts
$x_{t-1}$ and add some noise to it with the transition in order to get
$x_{t}$.

The reverse procedure, on the other hand, begins at $x_{t}$ and attempts to eliminate the noise that has been produced in an effort to try to retrieve $x_{t-1}$ during the transition. A sample from the dataset is transformed into pure noise by doing $T$ steps of the forward process with $t = 1, 2, ..., T$. The noise, on the other hand, becomes a sample of the input data distribution after $T$ steps of the reverse process are completed.
Diffusion models function by gradually adding noise to the data distribution via the forward process in $T$ stages, until pure Gaussian noise is obtained.
Afterwards, learn how to reverse the procedure to clean out the samples' noise.

To achieve sampling, it would be possible to start with noise and carry through T steps backwards to transform it into a data sample.
The lack of learnable parameters in the forward process is one of the unique features of DDPM. The Gaussian noise is added in $T$ stages using a fixed Markov chain in accordance with a predetermined noise schedule of $\beta$1, $\beta$2, ..., $\beta$T. Moreover, it has been demonstrated that the reverse process $p_{\theta}(x_{t-1}|x_t)$ is likewise a Gaussian if the forward process' transitions are Gaussian and $\beta$t is small. The number of steps T must be as high as feasible due to this. Due to the fact that the higher T, the smaller the $\beta$t will be, and the more true the preceding statement is.

\subsubsection{Progressive distillation diffusion}
\label{sssec:subsubhead}

Progressive distillation algorithm proposed by~\cite{DBLP:journals/corr/abs-2202-00512}. It is a novel parameterization of diffusion models that can improve their stability while utilizing fewer sampling steps, as well as a technique to distill a trained deterministic diffusion sampler into a new diffusion model that uses half as many sampling steps. The authors show that this progressive distillation method, which requires no more time than training the initial model, may be an effective solution for generative modeling employing diffusion at both train and test time. The technique intends to drastically shorten the sampling time of diffusion models in unconditional and class-conditional generation while maintaining good sample quality.
 
\begin{algorithm}[H]
\centering
\caption{Progressive distillation proposed by~\cite{DBLP:journals/corr/abs-2202-00512} }\label{alg:dif_distill}
\begin{algorithmic}
\Require {{Trained teacher model $\hat x_{\eta}(z_t)$}}
\Require Data set $\mathcal{D}$
\Require Loss weight function $w()$
\Require {{Student sampling steps $N$}}
\For{{{$K$ iterations}}}
\State {{$\theta \leftarrow \eta$}} \Comment{Initiate student from teacher}
\While{not converged}
\State $x \sim \mathcal{D}$
\State {{$t=i/N, ~~  i \sim Cat[1, 2, \ldots, N]$}}
\State $\epsilon \sim N(0, I)$
\State $z_t = \alpha_t x + \sigma_t \epsilon$
\State {{\texttt{\# 2 steps of DDIM with teacher}}}
\State {{$t' = t-0.5/N$,~~~ $t'' = t-1/N$}}
\State {{$z_{t'} = \alpha_{t'}\hat x_{\eta}(z_t) + \frac{\sigma_{t'}}{\sigma_{t}}(z_t - \alpha_t\hat x_{\eta}(z_t))$}}
\State {{$z_{t''} = \alpha_{t''}\hat x_{\eta}(z_{t'}) + \frac{\sigma_{t''}}{\sigma_{t'}}(z_{t'} - \alpha_{t'}\hat x_{\eta}(z_{t'}))$}}
\State {{$\tilde x = \frac{z_{t''}-(\sigma_{t''}/\sigma_{t})z_t}{\alpha_{t''}-(\sigma_{t''}/\sigma_{t})\alpha_t}$}} \Comment{Teacher $\hat x$ target}

\State $\lambda_t = \log[\alpha^{2}_t / \sigma^{2}_t]$
\State $L_{\theta} = w(\lambda_t) \lVert \tilde x -  \hat x_\theta(z_t)\rVert_{2}^{2}$
\State $\theta \leftarrow \theta - \gamma\nabla_{\theta}L_{\theta}$
\EndWhile
\State {{$\eta \leftarrow \theta$}} \Comment{Student becomes next teacher}
\State {{$N \leftarrow N/2$}} \Comment{Halve number of sampling steps}
\EndFor
\end{algorithmic}
\end{algorithm}

The instructor diffusion model, which is acquired by conventional learning, is the first step in the progressive distillation technique Algorithm~\ref{alg:dif_distill}.
Using the same parameters and the same model description, we initialize the student model with a copy of the instructor model at each iteration of progressive distillation.

We collect data from the training set, add noise to it, just as in traditional learning, and then we create learning losses by applying the learner model to denoise these noisy $z_t$ data. Progressive distillation differs fundamentally in that the objective for the denoising model is specified in a different way: rather than using the raw data x, we have the student models denoise to a target $\hat x$ where one student's DDIM step is equivalent to the instructor's two DDIM steps. Using the instructor, we do two DDIM sampling steps, beginning with $z_t$ and finishing with $z_{t-1/N}$, where $N$ is the total number of sample steps taken by the student.

The value that the student model should predict in order to progress from $z_t$ to $z_{t-1/N}$ in one step is then determined by inverting one DDIM step. With the teacher model and the beginning point $z_t$, the final goal value $\hat x (z_t)$ is completely specified, enabling the student model to make an accurate prediction when evaluating at $z_t$.

On the other hand, because of multiple diverse data points, the starting data point $x$ is not fully known at $z_t$. The first denoising model predicts a weighted average of the potential values of $x$, resulting in a fuzzy forecast, because $x$ can produce the same noisy data as $z_t$. The student model can go more quickly while sampling and can make clearer predictions.
We may repeat the process with $N/2$ steps by running the distillation to train the student model in $N$ sampling steps: The student model changes into the new instructor, and a copy of the original student model is used to initialize the new student model.

As opposed to how we learned our original models, we always run the progressive distillation in discrete time. We discretize this discrete time so that the largest time index corresponds to a signal-tone relation of zero, or $\alpha$1 = 0, which exactly matches the input noise distribution used for testing, $z_1 \tilde N (0, I)$.

\subsubsection{Parametrization and Training}
In this part, we go through how to provide the reconstruction loss weight $w(\lambda_t)$ and parameterize the denoising model $\hat x_\theta$. If $\sigma^{2}_t = 1 - \alpha^{2}_t$, a typical variance-preserving diffusion process is taken into consideration. This may be said without losing generality because other diffusion process specifications, such as the variance-exploding specification, can be viewed as comparable to it up to rescaling the noisy latents $z_t$. We employ the formula $\alpha_t = \cos(0.5\pi t)$.

The denoising model is often parameterized by directly predicting $\epsilon$ using a neural network $\hat\epsilon_{\theta}(z_t)$, which implicitly sets $\hat x_{\theta}(z_t) = \frac{1}{\alpha_t}(z_t - \sigma_t\hat\epsilon_{\theta}(z_t))$. The training loss in this instance is often specified as the mean squared error in the $\epsilon$-space:

\begin{equation}
\begin{split}
L_{\theta} = \lVert \epsilon - \hat\epsilon_{\theta}(z_t)\rVert_{2}^{2} = \\ \left\| \frac{1}{\sigma_t}(z_t - \alpha_tx) - \frac{1}{\sigma_t}(z_t - \alpha_t\hat x_{\theta}(z_t))\right\|_{2}^{2} = \\ \frac{\alpha^{2}_t}{\sigma^{2}_t} \lVert x - \hat x_{\theta}(z_t) \rVert_{2}^{2},
\end{split}
\label{eq:loss}
\end{equation}

which can thus equivalently be seen as a weighted reconstruction loss in $x$-space, where the weighting function is given by $w(\lambda_t)=\exp(\lambda_t)$, for log signal-to-noise ratio $\lambda_t = \log[\alpha^2_t/\sigma^{2}_t]$.

Although this standard specification is effective for training the original model, it is not suitable for distillation. This is because as progressive distillation advances, we increasingly evaluate the model at lower and lower signal-to-noise ratios $\alpha^2_t/\sigma^{2}_t$ than we do when training the original diffusion model. When $\hat x_{\theta}(z_t) = \frac{1}{\alpha_t}(z_t - \sigma_t\hat\epsilon_{\theta}(z_t))$ divides by $\alpha_t \rightarrow 0$, the influence of tiny changes in the neural network output on the implied prediction in $x$-space increases as the signal-to-noise ratio approaches zero.

This is not a major issue when taking many steps since the $z_t$ iterates are clipped, and subsequent updates may fix any errors. But, when the number of sample steps is decreased, it becomes more crucial. The input to the model is ultimately just pure noise $\epsilon$, which corresponds to a signal-to-noise ratio of zero, or $\alpha_t=0, \sigma_t=1$, if we reduce the process all the way down to a single sampling step. At this point, the relationship between the prediction of $\epsilon$ and the forecast of $x$ totally disintegrates: observations with $z_t = \epsilon$ are no longer indicative of $x$, and predictions with $\hat\epsilon_{\theta}(z_t)$ are no longer implicitly predictive of $x$. The weighting function $w(\lambda_t)$ assigns zero weight, as can be seen when looking at our reconstruction loss (equation \ref{eq:loss}).

In order for distillation to function, the diffusion model must be parameterized in a way that ensures the implicit prediction $\hat x_{\theta}(z_t)$ stays constant as $\lambda_t = \log[\alpha^2_t/\sigma^{2}_t]$ changes. All of the following alternatives that we explored seemed to complement progressive distillation well:
\begin{itemize}
    \item Predicting $x$ directly.
    \item Predicting both $x$ and $\epsilon$, via separate output channels $\{\tilde x_{\theta}(z_t), \tilde\epsilon_{\theta}(z_t)\}$ of the neural network, and then merging the predictions via $\hat x = \sigma^{2}_t\tilde x_{\theta}(z_t) + \alpha_{t}(z_t - \sigma_t\tilde\epsilon_{\theta}(z_t))$, thus smoothly interpolating between predicting $x$ directly and predicting via $\epsilon$.
    \item Predicting $v \equiv \alpha_t\epsilon - \sigma_{t} x$, which gives $\hat x = \alpha_t z_t - \sigma_t\hat v_{\theta}(z_t)$.
\end{itemize}

We must choose a suitable parameterization as well as a reconstruction loss weighting, or $w(\lambda_t)$. As the DDPM configuration inherently assigns a weight of zero to data with zero SNR and weights the reconstruction loss by the signal-to-noise ratio, it is not a good option for distillation. We examine two additional weightings for training loss:
\begin{itemize}   
    \item $L_{\theta} = \text{max}(\lVert x - \hat{x}_t \rVert_{2}^{2}, \lVert \epsilon - \hat{\epsilon}_t \rVert_{2}^{2}) = \text{max}(\frac{\alpha^{2}_t}{\sigma^{2}_t}, 1)\lVert x - \hat{x}_t \rVert_{2}^{2}$; ``truncated SNR'' weighting.
    \item $L_{\theta} = \lVert v_t - \hat{v}_t \rVert_{2}^{2} = (1+\frac{\alpha^{2}_t}{\sigma^{2}_t})\lVert x - \hat{x}_t \rVert_{2}^{2}$; ``SNR+1'' weighting.
\end{itemize}

As the sampling distribution used for $\alpha_t,\sigma_t$ has a significant impact on how much weight the predicted loss assigns to each signal-to-noise ratio, this sampling distribution must be taken into consideration when choosing a loss weighting in practice. Our findings apply to a cosine schedule with time uniformly sampled from $[0,\, 1]$ and $\alpha_t = \cos(0.5\pi t)$.
\subsection{U-Net}
\label{ssec:subhead}
 
\subsubsection{1D U-Net}
\label{sssec:subsubhead}

In this study, we have employed the advanced and efficient 1D U-Net architecture. By considering each frequency as a distinct channel, we leverage the power of one-dimensional convolution cores which outperform their 2D counterparts in terms of speed. This allows us to effectively apply the U-Net architecture to both signals and mel-spectrograms.
The U-Net-based deep neural network (DNN) that we employ features an encoder-decoder structure equipped with a bottleneck layer. To simplify the presentation, we refer to a group of convolutional layers as a bulk block. Each convolution layer within the bulk block applies one-dimensional convolutions using filters of various sizes.
We introduce batch normalization (BN) to restrict the acquisition of large activation levels, ensuring that the network remains stable and less prone to overfitting.
In the U-Net based DNN, we incorporate ResNet blocks, which help create a smoother loss terrain surface. This not only enhances the overall performance of the model but also simplifies the process of fine-tuning.

To handle negative values, we apply a rectified linear unity (ReLU) activation function after the output of each convolution layer. This activation function aids in removing negative values and promotes the learning of useful representations.
Within each bulk block of the UNet-based DNN, a max-pooling operation is performed. This operation plays a crucial role in reducing the dimensions of the input by a factor of 2, facilitating a more compact representation of the data as it flows from one bulk block to the next with decreased mass.
The encoder and decoder components of the network are interconnected, facilitated by skip connections. These skip connections allow us to combine the last layer of each bulk block in the encoder with the decoder part. This integration enhances the model's ability to identify input events by providing it with more information.

To ensure that the output component maintains the same input size, we employ a 2x upsampling technique. This is achieved using transposed convolution followed by ReLU activation. The use of a UNet-based DNN enables us to seamlessly integrate high-level and low-level functionalities in the encoder pipeline, thereby improving the accuracy of input signal event detection.
Overall, the utilization of the 1D U-Net architecture, coupled with its unique features such as skip connections and max-pooling, allows us to build a powerful and effective deep learning model capable of accurately analyzing and processing input data.

\section{Experimental setup}
\label{sec:pagestyle}

The many experiments, mostly pertaining to the hyperparameters of diffusion models, will be covered in this section. To learn more about the parameters of diffusion models, a model will be fitted to a learning set and various hyperparameter configurations will be compared.

\subsection{Dataset and training setup}
\label{ssec:subhead}
For the training process 2 large high-quality audio datasets were collected and sliced to Normalized data.
\begin{enumerate}
    \item \textbf{SoTr dataset}: a 120Gb audio dataset of licence-free soundtrack .mp3 files was transformed into around 300\,000 slices of 3-channel $128 \times 128$ mel-spectrograms, saved and processed as tensors.
    \item \textbf{PiTr dataset}: a 7Gb audio dataset of licence-free piano music .mp3 files, which was transformed into around 60\,000 slices of 3-channel $128 \times 128$ mel-spectrograms, also saved and processed as tensors.
\end{enumerate}

The main trick to maintain high resolution is rearrange transformation from 1-channel $128 \times 384$ to 3-channel $128 \times 128$ tensor mel-spectrogram slices via connecting similar frequencies together by a snake-like packing and making additional channels and later use inverse transforming to reverse the process and get long mel-spectrogram.

The implementations have been conducted using Python 3.6 as well as the Pytorch and Librosa python libraries.

The training process for 1M steps takes around 1 week on ZHORES~\cite{zacharov2019zhores} cluster (104 GPUs (Tesla V100-SXM2) with 16Gb of memory). 

DDPM parameters: Number of steps $T$: 1000; Number of $T_{\text{infer}} = T$; Beta schedule: cosine.

Training parameters: Optimizer: Adam with initial learning rate of $2 \times 10^{-5}$ and cosine annealing lr scheduler; Epochs: 1000 ; Batch size: 8 ; Loss function: $L_2$.

\subsection{Evaluation}
\label{ssec:subhead}

Generative model estimation is not an easy process. It can be challenging to formalize audio synthesis, in particular when the objective is to create perceptually accurate sound.
Comparing models using audio samples and gauging how well they perform classification tasks is a popular activity. We compare the performance of our models to a wide range of metrics that are often used in the literature, each of which indicates a different component of the model performance.

\subsubsection{Fr\'{e}chet Audio Distance (FAD)}
Unlike other audio evaluation metrics, FAD~\cite{Kilgour2019FrchetAD} compares embedding statistics obtained on a comprehensive evaluation set with embedding statistics created on a sizable set of clean music (such as the training set), as opposed to focusing on specific audio snippets.
As a result, FAD becomes a reference-free measure that may be used to evaluate a set of data without access to the ground truth reference audio.

We then compute multivariate Gaussians on both the evaluation set embeddings 
$\mathcal N_e(\mu_e, \,\Sigma_e)$
and the reference embeddings 
$\mathcal N_r(\mu_r, \,\Sigma_r)$.
In~\cite{dowson1982frechet} it is shown that the Fr\'{e}chet
distance~$F$ between two Gaussians $\mathcal  N_b$ and~$\mathcal N_e$~is
\begin{equation*}
F(\mathcal N_b,\,\mathcal N_e)
=
\left\|\mu_b-\mu_e\right\|^2+
\hbox{tr}
\left(
\Sigma_b+\Sigma_e-
2\sqrt{\Sigma_b\Sigma_e}
\right).
\end{equation*}

Smaller gaps between synthetic and actual data distributions are indicative of lower FAD. When it comes to computational effectiveness, congruence with human assessments, resilience against noise, and sensitivity to intra-class mode dropping, FAD performs admirably.

\subsubsection{Pitch Inception Score (PIS) and Instrument Inception Score (IIS)}
    The Inception Score (IS) is defined as the mean KL divergence between the conditional class probabilities $ p(y|x)$, and the marginal distribution $ p(y)$ using the predictions of an Inception classifier~\cite{SalimansGZCRCC16}:
    \begin{equation}
        \exp{\big (E_x[KL(p(y|x)||p(y))]\big )}
        \label{eq:iscore}
    \end{equation}
    
By training the Inception Net on the tasks of instrument and pitch classification from mel-spectrograms, similar to~\cite{Engel}, we adapt this measure to audio assessment. They are known correspondingly as the Pitch Inception Score (PIS) and the Instrument Inception Score (IIS). IS penalizes models whose examples belong to only a small subset of the potential classes as well as models whose examples cannot be confidently sorted into a single class.
A conditional label distribution $ p(y|x)$ with a low entropy should be present in spectrograms that contain meaningful items. In addition, we anticipate that the model would produce a variety of sounds, therefore the marginal  $\int p(y|x = G(z))dz$ should have a high degree of entropy.This metric is found to be useful for the evaluation of image models, correlating well with human judgment, although it is not sensible to over-fitting~\cite{barratt2018note}. 
    
\subsubsection{Pitch Kernel Inception Distance (PKID) and Instrument Kernel Inception Distance (IKID)}

According to~\cite{BinkowskiSAG18}, the Kernel Inception Distance (KID) calculates the differences between samples taken separately from generated and actual distributions. The Maximum Mean Discrepancy (MMD) between Inception representations, squared. The produced probability distribution $(P_g)$ is more similar to the true data distribution $(P_r)$ when the MMD is smaller. We use the squared MMD estimator from~\cite{GrettonBRSS12} between $m$ samples of $X \sim P_r$ and $n$ samples of $Y \sim P_g$ for any fixed characteristic kernel function $k$, defined as:
    \begin{equation}
            \begin{split}
                \text{MMD}^2 (X, Y) =
                     \quad \frac{1}{m(m-1)}\sum_{i\neq j}^{m}k(x_i, x_j)  + \\ \quad \frac{1}{n(n-1)}\sum_{i\neq j}^{n}k(y_i, y_j)  - \\ \quad \frac{2}{mn}\sum_{i=1}^{m} \sum_{j=1}^{n} k(x_i,y_j)
            \end{split}
    \end{equation}
    
    Here, we use an inverse multi-quadratic kernel (IMQ) $k(x,y) = 1 / \bigl(1+\|x-y\|^2 / 2 \gamma^2\bigr)$ with $\gamma^2 = 8$~\cite{Rustamov}, 
    as it has a heavier tail than a Gaussian kernel, hence, it is more sensitive to outliers. We borrow this metric from the Computer Vision literature and apply it to the audio domain.

\section{Results}
\label{sec:typestyle}

The quantitative evaluation for samples generated by the models are shown in Table \ref{tab::methods_metrics:4}.

\begin{table}[H]
    \caption{Unconditional models for SoTr and PiTr. Higher is better for PIS and IIS, lower is better for PKID, IKID and FAD}
    \label{tab::methods_metrics:4}
    \begin{center}
        \begin{tabular}{ l l l l l l}
          \toprule
           {Models} & {PIS}  & {IIS} & {PKID} & {IKID} & {FAD}  \\
          \midrule
          DDPM + \\ UNET $\vert$ SoTr & 10.3 & 2.6 & 0.013 & 0.173 & 4.55 \\
          DDPM + \\ UNET $\vert$ PiTr  & \bf{11.5} & 1.9 & \bf{0.011} & 0.166 & 4.17 \\
          PDD + \\ 1D UNET $\vert$ SoTr   & 9.5 & 2.9 & 0.021 & 0.101 & 2.71 \\
          PDD + \\ 1D UNET $\vert$ PiTr  & 9.3 & \bf{3.6} & 0.016 & \bf{0.063} & \bf{2.13} \\
          \bottomrule
        \end{tabular}
    \end{center}
\end{table}

The observations stemming from the analysis of the straightforward Denoising Diffusion Probabilistic Model (DDPM) reveal a discernible pattern of deteriorating results. However, when we delve into the realm of more sophisticated models such as the Progressive Distillation Diffusion Model in conjunction with the 1D U-Net architecture (PDD + 1D UNET), a noteworthy transformation occurs in the generated outcomes, aligning them quite closely with the characteristics of genuine data across various metrics.

What's more, an intriguing discovery arises when we consider the PiTr dataset, as it consistently yields superior results. This phenomenon can be attributed to the presence of a simpler and less intrusive representation of the mel-spectrogram commonly found in piano music, which steers the generator towards encapsulating the multifaceted variations in pitch. Notably, unofficial listening tests support this claim, showcasing that the PKID, IKID, FAD exhibit stronger alignment with perceived sound quality compared to their counterparts, PIS and IIS.

Shifting our focus to the SoTr dataset, characterized by low values of PIS and IIS, an intriguing trend surfaces. In this context, PIS and IIS appear to align more effectively with the subjective evaluation of audio quality, unlike their performance with the PiTr data. In the latter case, PIS and IIS fall short in accurately reflecting the model's inability to create distinct pitches and faithfully replicate the distinctive timbral properties of the training data. Curiously, despite this discrepancy, the model's PIS and IIS scores remain significantly elevated.

These insights shed light on a critical aspect - the inadequacy of inception models in providing an accurate assessment of the overall quality of generated content. The lack of resistance to the particular artifacts inherent in these representations renders the inception models susceptible to misjudging the true essence of the generated content. Furthermore, the presence of lossy representations introduces an additional layer of bias, as the compression process itself may potentially distort the quantitative evaluation.

\section{Limitation and future work}
\label{sec:majhead}

Even if the suggested strategies have previously produced positive outcomes, some advancements are yet attainable. Then, a fixed constant based on the diffusion step and no learning parameters is determined as the reverse process variance. Recent research, however, has demonstrated an intriguing boost in sample quality in computer vision applications by learning it instead.

Also, it was suggested that not all diffusion processes affect sample quality equally. To balance the relative importance of each diffusion phase over the whole diffusion process, it is suggested that the noise schedule be changed.
The sampling speed of diffusion models is one of their major drawbacks. In fact, the network must do numerous sequential forward passes in order to generate one sample. Compared to other deed generating models, which often just need one, this is costly.

The sample pace can be slowed down, according to several articles. It is possible to employ fewer sample diffusion steps than those required for training. The basic premise of both approaches is to match the training variance schedule with the smaller variance schedule used during sampling and to uniformly distribute $N$ diffusion steps amongst the $T$ steps utilized at training.
More time could be spent on improving the architecture that was used. More contemporary U-Net could be used in place of the current U-Net, or other models could be looked at to add diffusion to.

\section{Conclusion}
\label{sec:print}

This work offers a deep learning method for the waveform-domain audio task of unconditional generation. Denoising diffusion probabilistic and progressive distillation diffusion models have all been looked into as a possible deep generative model. After that, a self-collected dataset has been used to comprehensively compare the produced approaches. A suitable deep learning architecture has been discussed, and some tests with parameters have been carried out.
Overall, this study adds a new audio generating tool to the machine learning toolkit. The constructed model still has significant flaws, though, and there are a number of approaches to make it better.

\bibliographystyle{IEEEbib}
\bibliography{strings,refs}

\end{document}